\documentclass[journal=ancac3,manuscript=article,layout=twocolumn]{achemso}

\usepackage[version=3]{mhchem} % Formula subscripts using \ce{}
\usepackage{siunitx}
\usepackage{parskip}

\DeclareSIUnit\angstrom{\protect \text {Å}}

\author{Martin Werres}
\affiliation{Institute of Engineering Thermodynamics, German Aerospace Center (DLR), Wilhelm-Runge-Str. 10, 89081 Ulm, Germany}
\alsoaffiliation[HIU]{Helmholtz Institute Ulm (HIU), Helmholtzstr. 11, 89081 Ulm, Germany}
\author{Yaobin Xu}
\affiliation{Environmental Molecular Sciences Laboratory, Pacific Northwest National Laboratory, Richland, WA, 99354, USA}
\author{Hao Jia}
\affiliation[PNNL2]{Energy and Environment Directorate, Pacific Northwest National Laboratory, Richland, WA, 99354, USA}
\author{Chongmin Wang}
\affiliation{Environmental Molecular Sciences Laboratory, Pacific Northwest National Laboratory, Richland, WA, 99354, USA}
\author{Wu Xu}
\affiliation[PNNL2]{Energy and Environment Directorate, Pacific Northwest National Laboratory, Richland, WA, 99354, USA}
\author{Arnulf Latz}
\affiliation{Institute of Engineering Thermodynamics, German Aerospace Center (DLR), Wilhelm-Runge-Str. 10, 89081 Ulm, Germany}
\alsoaffiliation[HIU]{Helmholtz Institute Ulm (HIU), Helmholtzstr. 11, 89081 Ulm, Germany}
\alsoaffiliation[University Ulm]{Department of Electrochemistry, University of Ulm, Albert-Einstein-Allee 47, 89081 Ulm, Germany}
\author{Birger Horstmann}
\affiliation{Institute of Engineering Thermodynamics, German Aerospace Center (DLR), Wilhelm-Runge-Str. 10, 89081 Ulm, Germany}
\alsoaffiliation[HIU]{Helmholtz Institute Ulm (HIU), Helmholtzstr. 11, 89081 Ulm, Germany}
\alsoaffiliation[University Ulm]{Department of Electrochemistry, University of Ulm, Albert-Einstein-Allee 47, 89081 Ulm, Germany}
\email{birger.horstmann@dlr.de}

\let\oldmaketitle\maketitle
\let\maketitle\relax

\title
  {Origin of heterogeneous stripping of lithium in liquid electrolytes}%\footnote{A footnote for the title}


\begin{document}


\twocolumn[
\begin{@twocolumnfalse}
\oldmaketitle
\begin{abstract}
  Lithium metal batteries suffer from low cycle life. During discharge, parts of the lithium are not stripped reversibly and remain isolated from the current collector. This isolated lithium is trapped in the insulating remaining solid-electrolyte interphase (SEI) shell and contributes to the capacity loss. However, a fundamental understanding of why isolated lithium forms and how it can be mitigated is lacking. In this article, we perform a combined theoretical and experimental study to understand isolated lithium formation during stripping. We derive a thermodynamic consistent model of lithium dissolution and find that the interaction between lithium and SEI leads to locally preferred stripping and isolated lithium formation. Based on a cryogenic transmission electron microscopy (TEM) setup, we reveal that these local effects are particularly pronounced at kinks of lithium whiskers. We find that lithium stripping can be heterogeneous both on a nanoscale and on a larger scale. Cryo TEM observations confirm our theoretical prediction that isolated lithium occurs less at higher stripping current densities. The origin of isolated lithium lies in local effects, such as heterogeneous SEI, stress fields, or the geometric shape of the deposits. We conclude that in order to mitigate isolated lithium, a uniform lithium morphology during plating and a homogeneous SEI is indispensable.  
\end{abstract}
\end{@twocolumnfalse}
]

\clearpage
\section{Introduction}
Lithium metal anodes paired with liquid electrolytes have regained much attention in the search for next-generation high-energy-density batteries~\cite{Horstmann2021,Liu2019,Zheng2020,Wang2022,Lin2017,Hobold2021}. Despite early commercialization attempts until the late 1980s, safety concerns and low durability hinder the successful use of lithium metal anodes~\cite{BRANDT1994173}. Recently, there has been much progress in monitoring the lithium metal structure during cycling, and there is the consensus that controlling the metal surface during cycling is key for successfully deploying lithium metal anodes~\cite{Li2014,Guan2018,Zheng2020,Liu2020,He2021}. The low durability originates from the continuous growth of solid-electrolyte interphase (SEI)\cite{Horstmann2021} and the formation of isolated lithium, which is electronically disconnected from the current collector~\cite{Fang2019}. Both effects lead to capacity loss and are enhanced with increasingly irregular, non-planar, and high-surface-area structures of the lithium anode. Experiments have found that the non-planar structure arises from whiskers, growing with no apparent growth direction during charging~\cite{STEIGER2014112,Kushima2017,He2019,Xu2020,Becherer2022}. Lithium whiskers, often with small diameters, entangle each other to form mossy lithium~\cite{Kushima2017}. The growth process is not limited by electrolyte transport at battery-typical current densities~\cite{Bai2016,Bai2018,Rulev2019,Becherer2021}. During discharging, portions of lithium remain isolated in the SEI shell~\cite{Fang2019,Yoshimatsu_1988,Li506}. This isolated lithium can accumulate at the electrode, increasing the cell resistance because the ion path becomes tortuous~\cite{Chen2017,Xu2019_Stefanopoulou,Gunnarsdttir2020} or float in the electrolyte~\cite{Kushima2017} and possibly react at high voltage~\cite{Liu2021_Cui} and elevated temperatures~\cite{Chang2020}. A fundamental understanding of these nano- and microscale effects would significantly contribute to developing mitigation strategies for these structural inefficiencies. 

Isolated lithium was first observed by Yoshimatsu \textit{et al.} Scanning electron microscopy (SEM) experiments showed that particle-like structures remain at the tip of lithium ``needles'' after stripping~\cite{Yoshimatsu_1988}. Li \textit{et al.} revealed by comparing cryo transmission electron microscopy (TEM) to room-temperature (RT) TEM that RT TEM greatly interacts with the lithium structures, and that the needles observed by RT TEM are lithium whiskers~\cite{Li506}. Cryo TEM is particularly useful for resolving the atomic scales of lithium whiskers and the SEI~\cite{Xu2020,Xu2020_2,Li506,LI20182167,Zhang2022,Fang2019}. The dynamics of the dissolution process of a single lithium whisker were captured by Steiger \textit{et al.} with optical microscopy techniques~\cite{STEIGER2014112}. It was observed that during stripping, a droplet is first disconnected at the tip, and afterwards, the root of the whisker dissolves, while a thin line connects the droplet to the anode. The thin line is most likely the hollowed-out SEI shell, but due to optical microscopy, the small structures cannot be resolved, and the pictures are governed by diffraction. Steiger \textit{et al.} discuss the possibility that the remaining droplet is an insoluble SEI particle~\cite{STEIGER2014112}. However, optical microscopy cannot investigate this hypothesis which requires spectroscopic analysis and high-resolution images of the residual particle.    

Recently, new experimental findings were supported by theoretical works which tried to understand phenomena associated with isolated lithium formation~\cite{LI20182167,Tewari2020}. Li \textit{et al.} observed in cryo TEM experiments that SEI nanostructure can induce notches where lithium whiskers are pinched-off and isolated lithium forms. Li \textit{et al.} tried to understand this by simulating lithium dissolution with locally enhanced ionic conductivity. A large enhancement factor of more than $1000$ is needed to simulate notches. This large enhancement factor is in contradiction to experimental estimates of ionic conductivities of SEI compounds, which tend to vary only one order of magnitude~\cite{Lin2017b,Huang2018}. Tewari \textit{et al.} found that an increasing discharge current density lead to less isolated lithium formation~\cite{Tewari2020}. Tewari \textit{et al.} tried to understand this with an atomistic kinetic Monte Carlo model, where lithium self-diffusion at the solid interface, lithium dissolution, and ionic diffusion in the electrolyte were incorporated, while neglecting effects of the SEI or lithium electromigration in the electrolyte~\cite{Tewari2020}. This model, reproduces that an increasing discharge current density leads to less isolated lithium. However, their simulation lattice of 150x100 lattice sites, i.e., atoms, is smaller than the typical whisker diameter and the observed thickness of dead lithium structures of approximately $\SI{100}{nm}$, equivalent to $\sim$285 atoms with the lattice constant of $\SI{351}{pm}$~\cite{Li506,Xu2020,Nadler1959}. In this model, the formation of isolated lithium is a purely stochastic process and cannot predict systematic formation of isolated lithium. Thus, the model is not complete and further theoretical investigation of isolated lithium is necessary.  

We investigate the lithium stripping process in a combined experimental and theoretical approach. We focus on the dissolution dynamics of lithium during the stripping of lithium metal anodes and the origin of isolated lithium. We aim to answer the following key questions: (1) Why does isolated lithium form? (2) How do the electrochemical conditions influence the formation of isolated lithium? (3) How can we mitigate the formation of isolated lithium? 

On the theoretical side, we developed a generalized phase-field model for lithium stripping underneath a rigid SEI. The model comprises the interaction of lithium with the SEI and the influence of geometrical effects on the dissolution rate. Both contributions are known to influence the reaction rate~\cite{Shi2018,Shi12138,Boyle2020,Yuan2019,Yang2016}. With this, we study the stripping behavior of a single lithium whisker. The model, described in detail in the Methods section, reproduces the literature observations. On the experimental side, we conducted cryo TEM to investigate lithium at different stripping stages and electron energy loss spectroscopy (EELS) to get insights into the chemical species of the whisker and the covering SEI.

%The work is structured as follows: In Section~\ref{Section:Results} we present our model predictions compared to the existing literature and validate them against new found results. In Section~\ref{Section:Conclusion} we conclude our findings and propose how isolated lithium can be mitigated. Lastly, in Section~\ref{Section:Method} we introduce both our model in Section~\ref{Section:Model} and our experimental setup in Section~\ref{Section:Experimental_Setup}. 

\section{Results and Discussion}
\label{Section:Results}

In order to understand the heterogeneous stripping of lithium whiskers, a thorough understanding of the structure and chemical composition of the whiskers is necessary. Thus, we first present experimental results of the chemical composition of the whisker and the covering SEI shell. The composition of the whisker is under debate, and the results will further be used to validate assumptions for our model. Second, the model is applied to predict the dissolution dynamics of a single straight lithium whisker. The results are compared to the observations of Steiger \textit{et al.}~\cite{STEIGER2014112}. In the cryo TEM setup, the dynamics of the dissolution process cannot be captured. However, it is a powerful tool for resolving the structures. Third, we thus show the micro and nanoscale observations of lithium after different stages of stripping with cryo TEM. A particular focus lies on kinked regions of the whiskers. Here, we compare the observed structures to model results of a 3D extension of the presented model. Finally, we extend our model for locally different SEI compositions and compare the results of our model to the notches observed by Li \textit{et al.} and their model~\cite{Li506}.

\begin{figure}[t]
     \centering
     \includegraphics[width=0.48\textwidth]{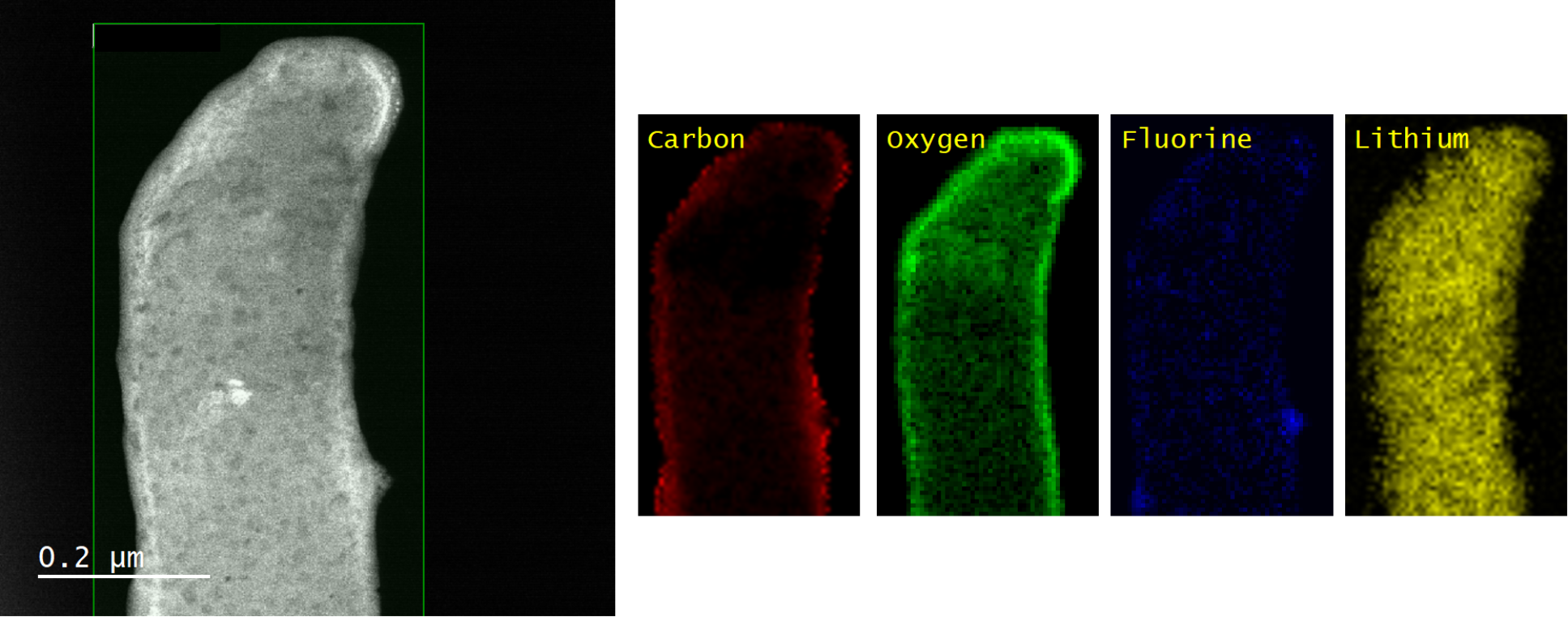}
     \caption{High-angle annular dark field (HAADF) scanning transmission electron microscopy image of the whisker and its tip with the corresponding electron energy loss spectroscopy (EELS) elemental mapping for a plating current density of $\SI{1}{A m^{-2}}$. Red color corresponds to carbon, green to oxygen, blue to fluorine, and yellow to lithium.}
     \label{Fig:EELS_low}
\end{figure}

\subsection{Chemical composition of the whisker and the SEI}

\begin{figure*}[t]
     \centering
     \includegraphics[width=0.95\textwidth]{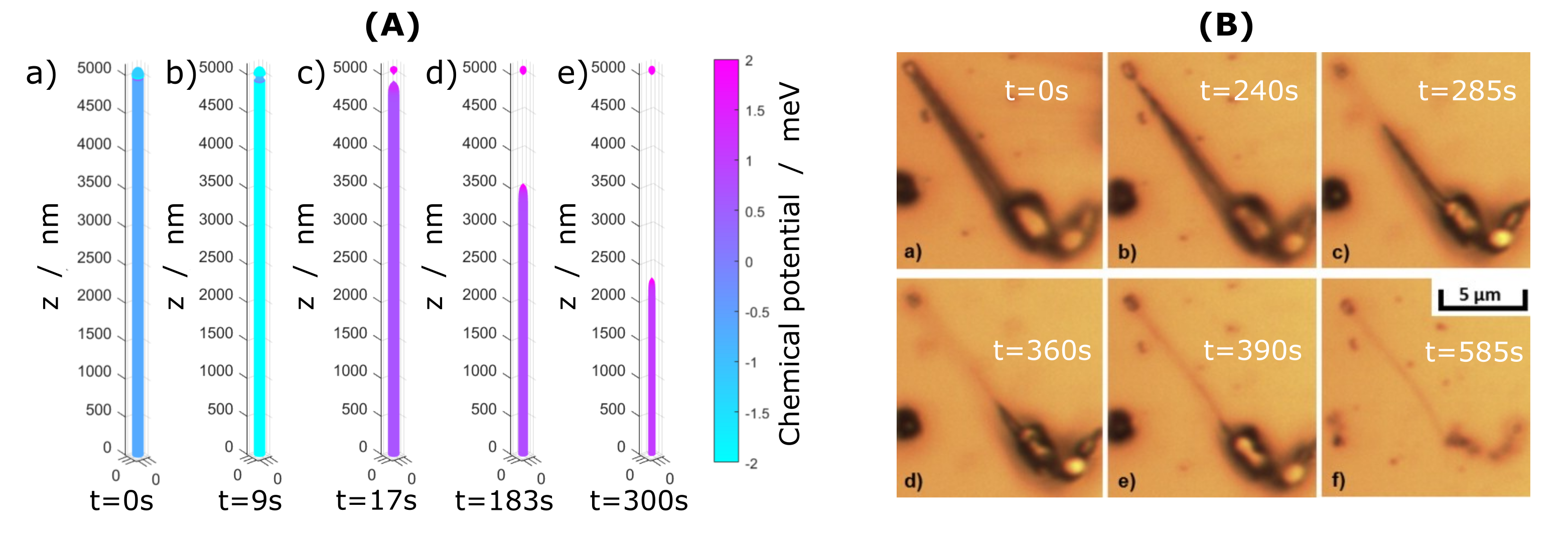}
     \caption{Comparison of simulated and observed whisker dissolution and droplet formation. (A) Snapshots of the simulation of the whisker dissolution at low current density $J=0.01 J_0$. The color represents the local chemical potential at the given time and location on the whisker surface. (B) Dissolution of a single lithium whisker as recorded in the experiment by Steiger et al.~\cite{STEIGER2014112}. Reprinted with the permission of Elsevier.}
     \label{Fig:comparison_steiger}
\end{figure*}

In the literature, the chemical composition of lithium whiskers, particularly of the tip, is still under debate~\cite{STEIGER2014112,Zachman2018,Fang2019,Xu2020_2,Xu2021_LiH,Shadike2021}. Steiger \textit{et al.} suggested that the whisker tip consists of an insoluble particle~\cite{STEIGER2014112}. 

For copper (Cu) whiskers occurring during electroplating, a similar phenomenon was experimentally verified~\cite{vanderMeulen1956}. There, a dirt particle induced much higher plating current densities at the tip, which led to whisker growth. However, lithium whiskers grow from the root, and different ideas for growth mechanisms are discussed in literature~\cite{Yamaki1998,Bai2016,Bai2018,Kushima2017,Yang2020,Rulev2020}. 

In order to clarify the chemical composition of the whisker tip, we perform EELS elemental mapping measurements of the whisker tip. The whiskers are formed by electrochemically depositing lithium on a Cu grid. As electrolyte, we use $\SI{1.2}{M}$ LiPF$_6$ in ethylene carbonate (EC)-ethyl methyl carbonate (EMC) (3:7 by wt) with 5 wt \% vinylene carbonate (VC) additive. As shown in Fig.~\ref{Fig:EELS_low}, we find that the whisker tip has the same chemical composition as the root of the whisker. It can be seen that the whisker has oxygen (O) and carbon (C) rich molecules and little fluorine (F) containing molecules. We associate this with the SEI-forming molecules. The intensity is higher at the shell where only SEI is imaged. In the core, the lithium intensity is higher, suggesting that the core is elemental lithium. The thickness of the shell, where C, O, and F are higher in intensity, is roughly $\SI{20}{nm}$. This matches with our measurement of the SEI in the cryo TEM images, as shown in Fig.~SI 4. Additionally, we see no strong dependence of the chemical composition of the SEI on the plating current density, as shown in Fig.~SI 2. Thus, the whisker tip is elemental lithium regardless of how the whisker was formed. There are no striking irregularities in the intensity distribution of the chemical compounds. We conclude that the SEI is homogeneous.

\subsection{Droplet formation during whisker dissolution}

With our model, we can simulate the galvanostatic dissolution of a single straight lithium whisker, taking the interaction between lithium and SEI into account. The SEI adheres to the lithium and lithium is stripped underneath the SEI. First, the adhesive bond breaks, influencing the local chemical potential, described by Equation~\ref{EQ:variational_chem_pot}, which in turn influences the local reaction rate, described by Equation~\ref{EQ:Butler_Volmer}. We apply our model to simulate the dissolution of a whisker at a low current density and compare our results with the recorded whisker dissolution dynamics of Steiger \textit{et al.}\cite{STEIGER2014112}. This experiment is ideal for our first comparison because the whisker is straight with no kinks and cylindrically symmetric.

As the exchange current density determines the dynamics for a given geometry and applied current, we state the applied current density $J$ relative to the exchange current density $J_0$. As discussed in the Methods section in more detail, the exchange current density depends on the used electrolyte and the thickness of the SEI, and is estimated to be $J_0=\SI{100}{A/m^2}$ based on experimental estimations~\cite{Shi12138,Chen2020}. We define our low current density scenario by $J=0.01 J_0$. In this case, the simulation results are shown in Fig.~\ref{Fig:comparison_steiger} (A). Depicted is the shape of the lithium whisker at different points in time. The three-dimensional curve describes the surface of the lithium metal at a given time and the surface color describes the local chemical potential of lithium at the surface.

The simulation predicts the nucleation of an instability just below the tip in the early stripping process, depicted in Fig.~\ref{Fig:comparison_steiger} (A)b). This point on the whisker surface is unique: there, the surface is concave opposed to everywhere else. The binding to the SEI is represented by a negative effective interfacial tension. Usually, convex surfaces dissolve preferentially but here, with the binding to the SEI, concave surfaces are preferred. The instability can be understood by the following: When the dissolution begins, the detachment of the SEI exposes new surface area of lithium. In general, the exposed surface is minimized during the dissolution process in order to minimize the surface energy. If the dissolution is slow, lithium is stripped at preferential places, leaving most of the lithium surface attached to the SEI. This leads to lithium being electronically disconnected from the current collector at the tip of the whisker, see Fig.~\ref{Fig:comparison_steiger} (A)c). After this,  at the tip, the remaining lithium forms a sphere, while the root of the whisker dissolves without any new instabilities, see Fig.~\ref{Fig:comparison_steiger} (A)d)-e). The part of the whisker which is still connected to the current collector then dissolves entirely.

The predicted dissolution behavior agrees nicely with the experimentally observed stripping of a lithium whisker by Steiger \textit{et al.}~\cite{STEIGER2014112}, see Fig.~\ref{Fig:comparison_steiger} (B). It was observed that below the tip, the whisker is thinned, and the tip gets disconnected from the root, as shown in Fig.~\ref{Fig:comparison_steiger} (B)b). Then, the root of the whisker dissolves completely, as shown in Fig.~\ref{Fig:comparison_steiger} (B)c)-e). 

\begin{figure}[t]
     \centering
     \includegraphics[width=0.48\textwidth]{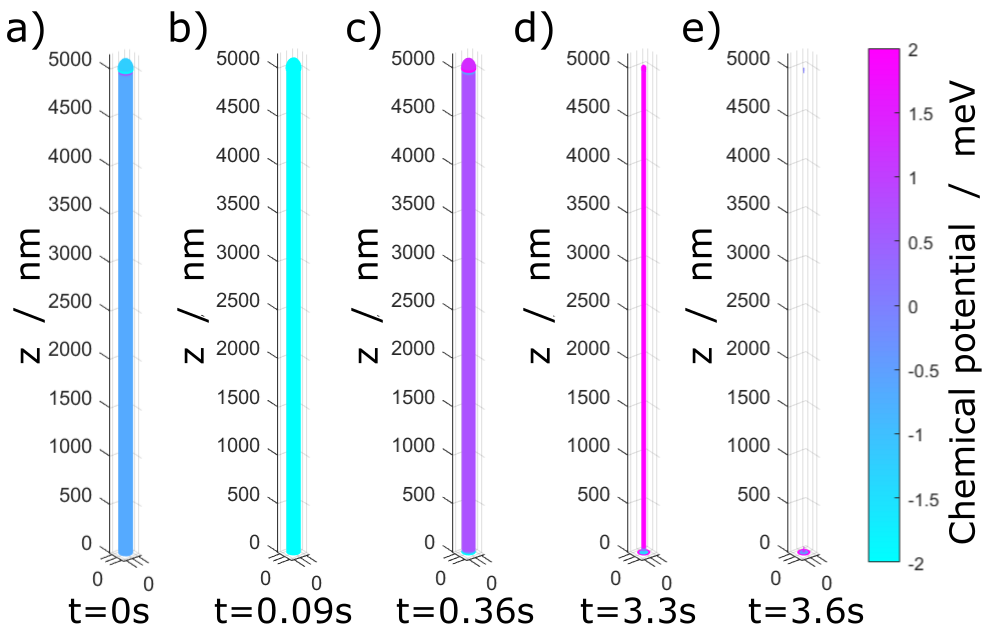}
     \caption{Snapshots of the simulation of the whisker dissolution at high current density $J=J_0$. The color represents the local chemical potential at the given time and location on the whisker surface.}
     \label{Fig:100_percent_with_SEI}
\end{figure}

Note that for the experiment by Steiger \textit{et al.}, a stripping current density of approximately $\SI{2}{\mu A cm^{-2}}$ was applied, estimated by dividing the total stripping current by the substrate area. The time scale of around $\SI{500}{s}$, over which the dissolution of the single whisker was observed, hints that the local dissolution current density of the whisker deviates from the reported global $\SI{2}{\mu A cm^{-2}}$. By approximating the whisker as a cylindrical object with a length of $\SI{10}{\mu m}$ and a diameter of $\SI{200}{nm}$, we estimate the dissolution current density of the whisker to be in the order of 
\begin{equation}
\begin{aligned}
    J&= I / \bar{A}_\text{whisker} = \frac{(F \cdot V_\text{whisker}) / (V_\text{M} \cdot t)}{\bar{A}_\text{whisker}} \\
    &\approx \SI{100}{\mu A cm^{-2}} = 0.01 J_0,
\end{aligned}
\end{equation}
with the Faraday constant $F$, the whisker volume $V_\text{whisker}$, the molar volume of lithium $V_\text{M}$, and the average whisker surface area during dissolution $\bar{A}_\text{whisker}$. Our estimation deviates from the reported average current density by two orders of magnitude. With our assumption of $J_0=\SI{100}{A/m^2}$, this estimation of the dissolution current density fits perfectly to our simulations with $J=0.01 J_0$.

Further, one can observe that in Fig.~\ref{Fig:comparison_steiger} (B) between a) and b) (in $\SI{240}{s}$), and between b) and c) ($\SI{45}{s}$), a comparable amount of lithium is stripped in different time intervals. We interpret this to imply that an onset time exists before the dissolution of the whisker starts. Our observations, discussed in the section below, support this interpretation. The underlying course of the existence of the onset time is not understood. We suggest that heterogeneous current distribution or heterogeneous kinetic barriers due to fluctuating surface properties of the individual whiskers can induce the onset time. The latter is well described in the context of our model. We predict locally varying dissolution currents depending on the local chemical potential, which in turn depends on the SEI properties and the whisker surface properties. Considering multiple whiskers with varying radii and heterogeneous SEI coverage, our local dissolution currents can lead to whiskers dissolving one at a time. We do not have an onset time in our simulation, as we consider only a single whisker. Thus, we predict the disconnection of the isolated lithium droplet after just a few seconds, as shown in Fig.~\ref{Fig:comparison_steiger} (A)c).

The discussed instability below the tip vanishes for higher stripping current densities, as depicted in Fig~\ref{Fig:100_percent_with_SEI}. For this simulation, we choose the stripping current density to be $J=J_0$. In this scenario, the local variations of interfacial tension become irrelevant and the local stripping current density variations are negligible~\cite{Horstmann2013,Kolzenberg2022,Bazant2017,Fraggedakis2020}. Therefore, the lithium-SEI bond breaking occurs homogeneously, see Fig.~\ref{Fig:100_percent_with_SEI} b) and c). During stripping, the whisker is thinned until it is dissolved, as depicted in Fig.~\ref{Fig:100_percent_with_SEI} d)-e). The remaining isolated lithium is hardly visible, Fig.~\ref{Fig:100_percent_with_SEI} e). This agrees with the experimental observations of Tewari \textit{et al.} of less isolated lithium formation at higher stripping current densities~\cite{Tewari2020}.

To validate the simulation results, we perform an additional experiment at a higher stripping current density of $\SI{10}{A m^{-2}}$, depicted in the Supporting Information (SI). After discharge, the whiskers are stripped completely and only a hollowed-out SEI shell remains, see Fig.~SI 3. This agrees with our theoretical predictions.

\begin{figure}[t]
     \centering
     \includegraphics[width=0.45\textwidth]{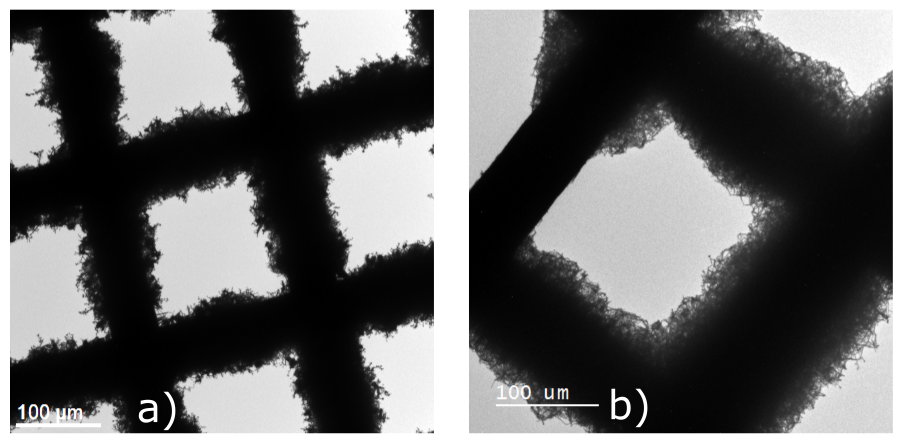}
     \caption{Cryo TEM image of the copper grid a) after 100 minutes plating at $\SI{1}{A m^{-2}}$ and b) after 500 minutes stripping at $\SI{0.1}{A m^{-2}}$. After non-uniform stripping, there are areas with high and low remaining whisker densities.}
     \label{Fig:microscale_plating_stripping_low}
\end{figure}

\subsection{Micro- and nanoscale observation of the stripping heterogeneity}

\begin{figure*}[t]
     \centering
     \includegraphics[width=0.95\textwidth]{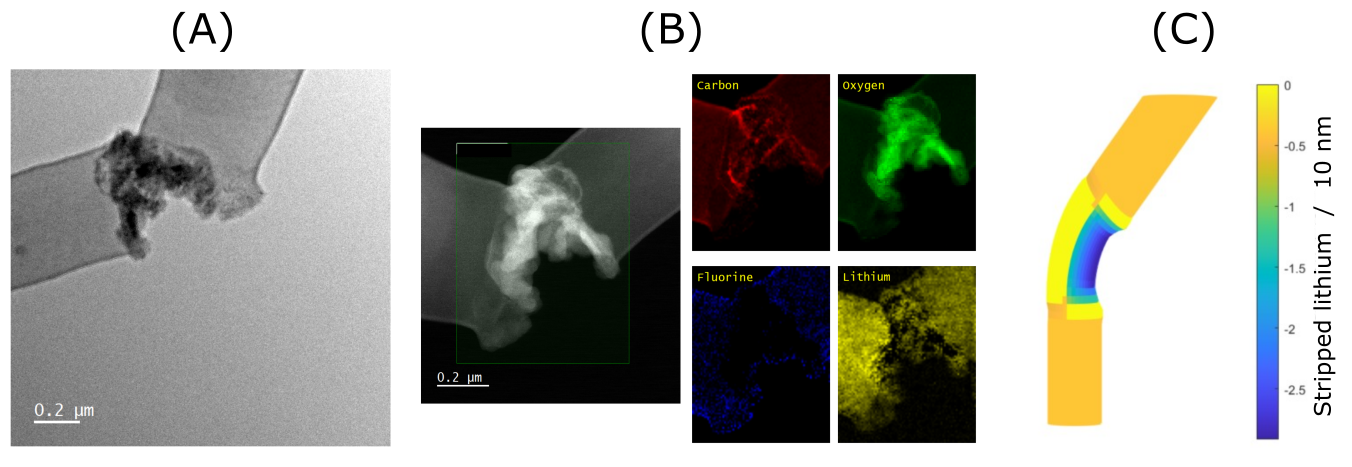}
     \caption{Stripping of kinked region. (A) Cryo TEM image of a whisker kink region after incomplete stripping at $\SI{0.1}{A m^{-2}}$. (B) HAADF STEM image of the whisker kink region after incomplete stripping at $\SI{0.1}{A m^{-2}}$ and the corresponding electron energy loss spectroscopy elemental mapping. Red color corresponds to carbon, green to oxygen, blue to fluorine, and yellow to lithium. (C) Snapshot of the simulation of whisker dissolution at $J=0.01J_0$ with focus on the whisker kink region at $t=\SI{3}{s}$. The color represents the amount of lithium stripped at the given time point compared to the original whisker surface.}
     \label{Fig:kink_stripping_low}
\end{figure*}

In order to understand the stripping behavior of lithium, one has to understand the dynamics at different length scales simultaneously, which is a very challenging task. At the centimeter scale, non-uniform distribution of current density is observed~\cite{Yari2022}. In our cryo TEM setup, we can probe for heterogeneous stripping ranging from several hundred microns down to a few nanometers. In order to investigate irregularities of lithium stripping in the length scale of $\SI{100}{\mu m}$, we take images of the copper grid after plating and after stripping. To investigate smaller length scales, we focus on small, interesting parts of the lithium whiskers after stripping.

In the case of stripping at a low current density of $\SI{0.1}{A m^{-2}}$, we observe non-uniform lithium dissolution on the microscale, as shown in Fig.~\ref{Fig:microscale_plating_stripping_low}. In the experiment, we stripped away about half of the plated lithium. Opposing to uniform stripping of the lithium whiskers, we observe areas where the lithium seems to be almost completely dissolved, while in other areas, it seems that dissolution has not started at all. From this, we conclude that larger-scale heterogeneity plays a critical role in the preferential dissolution of lithium. Local stress fields or locally different SEI compositions can cause this heterogeneity and lead to locally different overpotential and thus enhance or retard localized stripping. 

As we want to investigate the lithium whiskers during and after stripping, we focus on the areas with low remaining whisker density. There, the dissolution is mostly complete, and we can investigate if isolated lithium forms. We find that preferential stripping occurs mostly at kinks. In Fig.~\ref{Fig:kink_stripping_low} (A), we show a typical image of the observed structure at kinks. It can be seen that the preferential dissolution at the kinks leads to a separation point where one part of the whisker is electronically disconnected from the other part and thus forms isolated lithium. The connection remains only through a SEI, which seems to be different compared to the rest of the whisker covering SEI. This can be caused by two effects: First, by mechanical deformation of the native SEI covering the whisker, and second, by new reactions of exposed lithium with the electrolyte. In order to understand the cryo TEM image, we performed EELS elemental mapping to analyze the SEI composition, see Fig.~\ref{Fig:kink_stripping_low} (B). We can identify an O-rich SEI, with little amounts of C and very little amount of F. This suggests the formation of Li$_2$O nano-particles at the kink.

Kink regions have the distinct feature of different surface curvatures on the inside and the outside of the kink. Following our line of argument presented above, this geometry effect should lead to different stripping rates in the kink region at low current rates. We extend the whisker model to three dimensions to study if our predictions match the experimental observations. For this, we trace points on the lithium surface and calculate the surface curvature utilizing differential geometry as the eigenvalues of the shape operator. Fig.~\ref{Fig:kink_stripping_low} (C) shows the amount of stripped lithium in the vicinity of a kink in the early stages of stripping for a low stripping current density. It can be seen that stripping occurs preferentially on the inside of the kink, where the kink surface was initially concave. The blue color indicates that a large amount of lithium is stripped away, while the yellow color indicates that almost no lithium is stripped. In the blue region on the inside of the kink, the lithium-SEI bond is broken first. With further stripping, the preferred dissolution at the kink leads to a pinch-off of the upper whisker part and isolated lithium. After the pinch-off and the breakdown of the SEI shell, the isolated lithium part is fragile, can rotate, and potentially mechanically disconnect from the root of the whisker. Our predictions agree with our cryo TEM observations and explain why kinks are prone to isolated lithium formation. This explains why mossy lithium is particularly bad and why stripping behavior is better when the whiskers are straight and aligned~\cite{Fang2021}.

Note that our model considers a local chemical potential due to the lithium-SEI interaction through an adhesive bond. Additionally, the SEI can put pressure on lithium during whisker growth when lithium is plated underneath the SEI. Differences in the local stress distribution also cause differences in the local chemical potential and thus lead to different local stripping behavior. In kinks, the rotational symmetry of the geometry is broken. We thus anticipate that local stress fields can also play a role in the preferred stripping behavior of kinks. Investigating this effect would require extending our model with a model for the local stress distribution. This is possible because of the generality of our framework but outside the scope of this work. 

\begin{figure*}[t]
     \centering
     \includegraphics[width=0.95\textwidth]{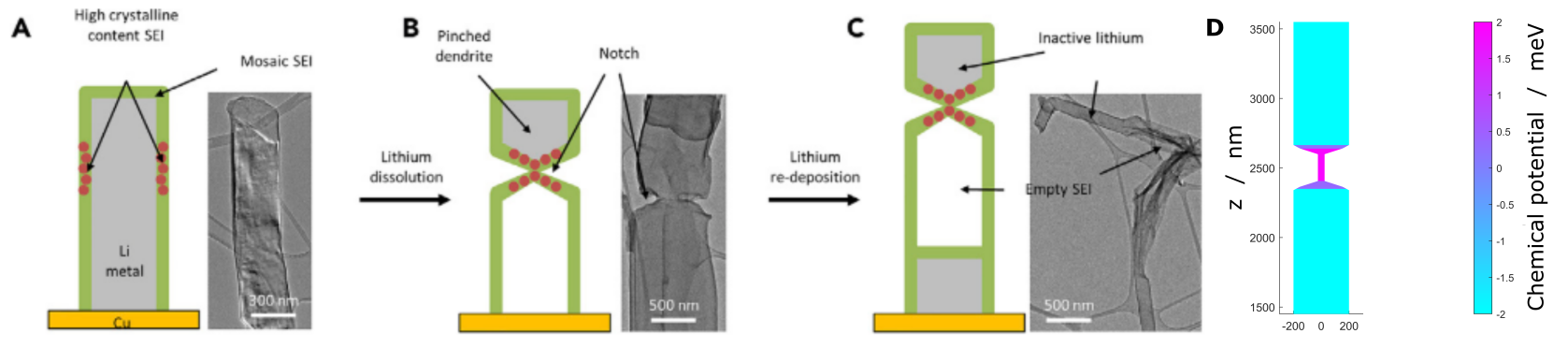}
     \caption{Comparison of notches as observed in experiments and predicted by simulation. A-C: Notches as observed by Li $\textit{et al.}$\cite{Li506} Places with locally different SEI nanostructure have higher ionic conductivity, and isolated lithium forms. Reproduced with permission by Elsevier\cite{LI20182167}. D: Snapshot of a simulation of Li whisker with enhanced exchange current density $S=1.5$ at a current density of $J=0.1 J_0$. A notch forms at the place with enhanced SEI properties.}
     \label{Fig:Li_notches}
\end{figure*}

\subsection{Notches}
Li \textit{et al.} found that during stripping notches can occur, as depicted in Fig.~\ref{Fig:Li_notches} A-C ~\cite{Li506}. Notches are seemingly random spots in the whisker where a part of the whisker is completely pinched off, leaving isolated lithium disconnected from the current collector. Li \textit{et al.} suspect that notches emerge at spots where the covering SEI has a higher ionic conductivity. Possibly through a locally different SEI composition. We adopt this idea and translate the locally enhanced ionic conductivity to a locally enhanced exchange current density with enhancement factor $S$. The exchange current density strongly depends on the SEI composition and thickness~\cite{Shi12138,Chen2020}. We explore what our model predicts by introducing a $\SI{90}{nm}$ long spot where the exchange current density is increased by $S=2$. The simulation results are depicted in Fig.~\ref{Fig:Li_notches} D. The results look remarkably similar to the observation of Li \textit{et al}. The enhanced exchange current density is equivalent to the enhanced ionic conductivity, but our model introduces an additional mechanism to the process. Due to the locally accelerated dissolution, the bond between lithium and SEI is broken faster. Thus, the reaction is even more enhanced, and notches can develop.

\begin{figure}[t]
     \centering
     \includegraphics[width=0.45\textwidth]{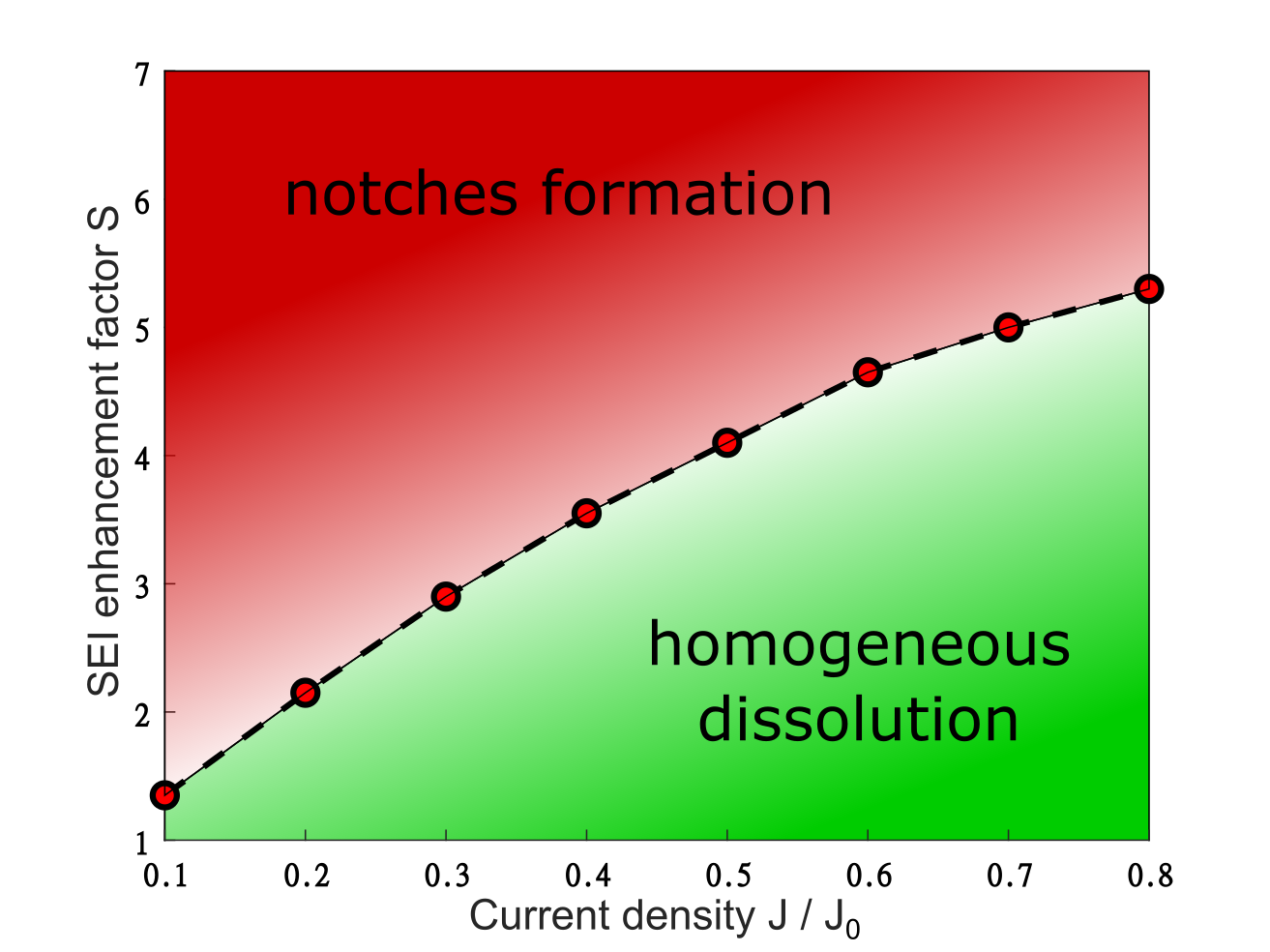}
     \caption{Phase-diagram of the stability of whiskers against notches as a function of applied current density and the enhancement factor $S$ of the locally higher conductive SEI. The red dots represent the minimum enhancement factor $S_\text{min}$ for notches to develop before other the lithium-SEI bond is broken at the rest of the whisker. The red area describes where notches are anticipated and the green area where notches are mitigated.}
     \label{Fig:stability_notches}
\end{figure}

In order to understand the influence of heterogeneity of the SEI on the formation of notches, we conduct a parameter study to find the minimum factor $S_\text{min}$ for notches to develop. We find that $S_\text{min}$ is larger for higher stripping currents, see Fig.~\ref{Fig:stability_notches}. We note that our model predicts trends, not quantitative values, as the exact value of $S_\text{min}$ depends on many factors, e.g. the whisker thickness or the surface area of the ionically higher conductive SEI. The value range of $S_\text{min} \sim 2-5$ for notches to occur seems more realistic than an enhanced ionic conductivity factor of 1000, as reported in the studies of Li \textit{et al.}\cite{LI20182167} We therefore conclude that the Li-SEI interaction is important for the occurrence of notches.

We propose the following interpretation of our results: 1. Our model confirms the idea from Li \textit{et al.} that a locally enhanced ionic conductivity of the SEI can lead to notches. 2. For smaller stripping current densities, notches can more easily occur and are thus more likely to occur. As discussed above, local variations play a lesser role at higher stripping current densities~\cite{Horstmann2013,Kolzenberg2022,Bazant2017,Fraggedakis2020}. Thus higher stripping current densities can mitigate the emergence of notches and thus can lead to less isolated lithium.

In our experiments, we do not observe notches, which we attribute to a more homogeneous SEI structure and composition.

\section{Conclusion}

 %For stripping a single lithium whisker at low current density, our model predicts the formation of a lithium droplet at the tip of the whisker. At higher discharge current densities, the droplet at the tip is not predicted to form, in line with the observation of less isolated lithium at higher current densities~\cite{Tewari2020}. When assuming a locally enhanced ionic conductivity in the covering SEI, the model predicts notches in the whiskers during stripping.

% We find that for small current densities, the whiskers dissolve heterogeneously. For higher current densities, the whiskers are mostly fully dissolved, leaving behind a hollowed-out SEI shell. The stripping process is non-uniform on both the micro- and nanoscale. On the microscale, it was found that stripping is heterogeneous, meaning that there are preferential stripping sites at the current collector where the stripping process initiates first, while other places start to dissolve much later. At the nanoscale, there are characteristic places at whiskers where stripping seems to be favored, kinks and the tip. After stripping, the SEI shows a different structure at those characteristic places than after plating.

\label{Section:Conclusion}
We investigated the stripping behavior of lithium at low stripping current densities and the origins of its heterogeneity in a combined theoretical and experimental approach. 
On the theoretical side, we developed a model for lithium whisker dissolution. Lithium whiskers occur in the early stages of electroplating and lead to mossy lithium. Our model comprises the interaction between lithium and SEI which is crucial to describe the experimentally observed phenomena. We predicted the occurrence of isolated lithium emerging at geometrically distinct spots below the tip or at kinks at low stripping current densities. The dissolution dynamics can describe the experimental observation by Steiger \textit{et al.}~\cite{STEIGER2014112} and the formation of a lithium droplet at the whisker tip. The model also reproduces the notches that can lead to isolated lithium, reported by Li \textit{et al}~\cite{Li506}.
On the experimental side, we plated lithium on a Cu TEM grid and investigated the emerging structures with cryo TEM, as this preserves the native state of the specimen~\cite{Xu2020,Xu2020_2,Zachman2018,Li506,Huang2020,LI20182167}. We observed that kink regions are very prone for isolated lithium formation. With EELS elemental mapping, we observed the SEI composition and found that the SEI composition changes, where isolated lithium is formed. We do not only observe that lithium stripping is non-uniform on a nanoscale of $\sim \SI{100}{nm}$ but also on a micro scale of $\sim \SI{100}{\mu m}$. We thus conclude that defects play a critcial role in the dissolution of lithium and that local effects, such as stress fields or local overpotential, can retard or facilitate lithium stripping.
From our study we can conclude that in order to prevent non-uniform stripping, morphological inefficiencies like whiskers or locally different SEI should be mitigated. Otherwise isolated lithium and low Coulombic efficiency are anticipated, especially at low discharge current densities. There, the observed non-uniform stripping phenomena are more prominent compared to higher stripping current densities.

\section{Methods}
\label{Section:Methods}

\subsection{Model}
\label{Section:Model}

We model the dissolution of a single lithium whisker without kinks, covered by an SEI layer of uniform thickness, during galvanostatic stripping as a reaction-limited process. Lithium whiskers are thought to be formed in a stress relaxation mechanism \cite{Yamaki1998,Kushima2017,Wang2018}. We assume that all stresses are relaxed with the start of the stripping and that the dissolution can be described solely by electrochemical reactions. We assume the diffusion of lithium in the electrolyte to be sufficiently fast and a constant concentration of Li$^+$ at the surface, which is valid for stripping current densities before diffusion limitations occur (see SI Eq.~A1\cite{Valoen2005}):
\begin{equation}
    j\ll j_\text{diff}\approx\SI{300}{A~m^{-2}}.
\end{equation}

\begin{figure}[t]
     \centering
     \includegraphics[width=0.48\textwidth]{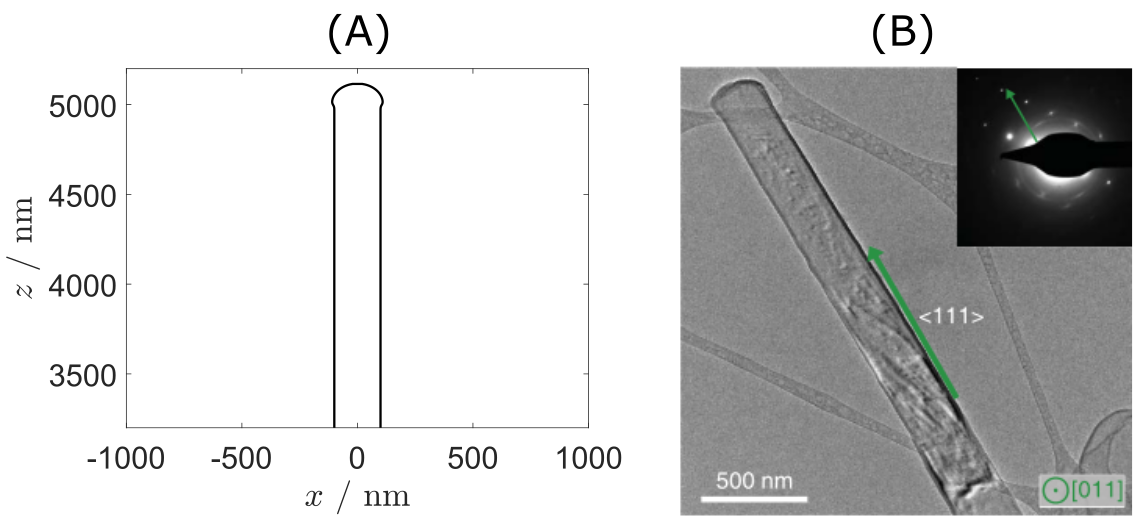}
    \caption{Comparison of the whisker geometry as assumed in the model and as seen in experiments. (A) Upper part of the whisker as described by our model assumptions. The body has a cylinder shape with a spherical tip that has a slightly bigger radius than the body. (B) Cryo TEM image of the upper part of a lithium whisker~\cite{Li506}. The tip is rounded, slightly spherical-like, and appears to be slightly bigger in radius than the whisker body. Reprinted with permission by The American Association for the Advancement of Science.}
    \label{Fig:geometry}
\end{figure}

As the initial geometry, we assume a cylinder-like shape with radius $R=\SI{100}{nm}$ and a spherical tip, as described in SI Eq.~A2. This resembles the structures observed in experiments, see Fig.~\ref{Fig:geometry}. This ansatz allows us to use a cylindrical symmetry. We describe the surface of the whisker by 
\begin{equation}
\label{EQ:xi_parametrization}
    (\xi,\phi) \mapsto (r(\xi),z(\xi),\phi)
\end{equation}
with the surface marker $\xi$. Here, $(r_0(\xi),z_0(\xi))$ is the initial whisker surface and the position of the SEI, which we assume to be rigid, i.e., it does not change during dissolution. In reality, the SEI shell falls together due to a negative pressure beneath the SEI surface. This is not important for our simulation of the whisker dissolution, as this happens after the lithium-SEI bond is broken.

Unlike canonical phase-field models, where the solid/liquid phases are captured by a phase parameter and a diffuse edge~\cite{Ely2014,Zhang2021_phase_field,Cogswell2015,Hong2018,Chen2015,Mu2019,Single2016,Castelli2021}, we assume a sharp edge and only track the surface of the whisker~\cite{Horstmann2013,Kolzenberg2022}. The dynamics of the whisker dissolution can then be described by the propagation of the surface points $\xi$ which move perpendicular to the surface:
\begin{align}
\label{EQ:equation_of_motion_1}
    \frac{\partial r(\xi)}{\partial t} = \frac{\Dot{z}}{\sqrt{\Dot{r}^2+\Dot{z}^2}} \cdot \frac{V_\text{M}}{F} \cdot J(\xi) \ , \\
    \label{EQ:equation_of_motion_2}
    \frac{\partial z(\xi)}{\partial t} = \frac{-\Dot{r}}{\sqrt{\Dot{r}^2+\Dot{z}^2}} \cdot \frac{V_\text{M}}{F} \cdot J(\xi) \ .
\end{align}
Here, $\Dot{r}=\partial r / \partial \xi$, $\Dot{z}=\partial z / \partial \xi$, $V_\text{M}=\SI{13.02e-6}{m^{3}/\text{mol}}$ is the molar volume of lithium, $F=\SI{96485}{C/mol}$ is the Faraday constant and $J(\xi)$ is the electrical current density, given by the Butler-Volmer expression
\begin{equation}
\label{EQ:Butler_Volmer}
    J(\xi)=J_0 \left[ e^{\frac{-F \Delta \Phi}{2RT}} - e^\frac{\mu(\xi)}{RT} e^{\frac{F \Delta \Phi}{2RT}} \right] , 
\end{equation}

with the effective exchange current density $J_0$, the ideal gas constant $R=\SI{8.314}{J/(mol ~ K)}$, the room temperature $T=\SI{298.15}{K}$, the potential step $\Delta \Phi=\Phi - \Phi_0$ relative to the lithium metal and the chemical potential $\mu(\xi)$ of lithium at the whisker surface~\cite{Horstmann2013,Kolzenberg2022}. Note that with the simplified formula for Marcus-Hush-Chidsey kinetics by Bazant and co-workers~\cite{Zeng2014}, equation \ref{EQ:Butler_Volmer} can be modified to better describe behavior for high overpotential~\cite{Boyle2020}, see SI Eq.~A5. The effective exchange current density depends on the used electrolyte and the thickness of the SEI. As this quantity is hard to measure we assume $J_0=\SI{100}{A/m^2}$ which is the reported order of magnitude for the exchange current density~\cite{Shi12138}. To avoid a dependency of our results on the value of $J_0$ we later give the initial stripping current density relative to the effective exchange current density. The chemical potential $\mu$ determines the non-equilibrium thermodynamics and follows from the Gibbs free energy
\begin{equation}
\label{EQ:Gibbs_free_energy}
\begin{aligned}
    G&=\int g \text{d}z=\int \sigma \text{d}A \\ &= \int \sigma(d,\alpha) 2 \pi r  \sqrt{1+{r}^{\prime 2}} \text{d} z
\end{aligned}
\end{equation}
which is based on the interfacial tension $\sigma(d,\alpha)$, with $r^\prime=\text{d}r/\text{d}z$. From equation \ref{EQ:Gibbs_free_energy} we get an expression for the free energy density $g$ which we can use to calculate the chemical potential via a variational derivative
\begin{equation}
\label{EQ:variational_chem_pot}
\begin{aligned}
    \mu&=\frac{\delta G[n]}{\delta n}=\frac{V_\text{M}}{2 \pi r } \frac{\delta G[r]}{\delta r} \\ &= \frac{V_\text{M}}{2 \pi r } \left(  \frac{\partial g }{\partial r} - \frac{\text{d}}{\text{d} z} \frac{\partial g }{\partial \frac{\partial r}{\partial z}} \right)
\end{aligned}
\end{equation}
where $n(z)=\pi r^2 / V_\text{M}$ is the number of lithium atoms per $z$-interval.

\begin{figure}[t]
     \centering
     \includegraphics[width=0.45\textwidth]{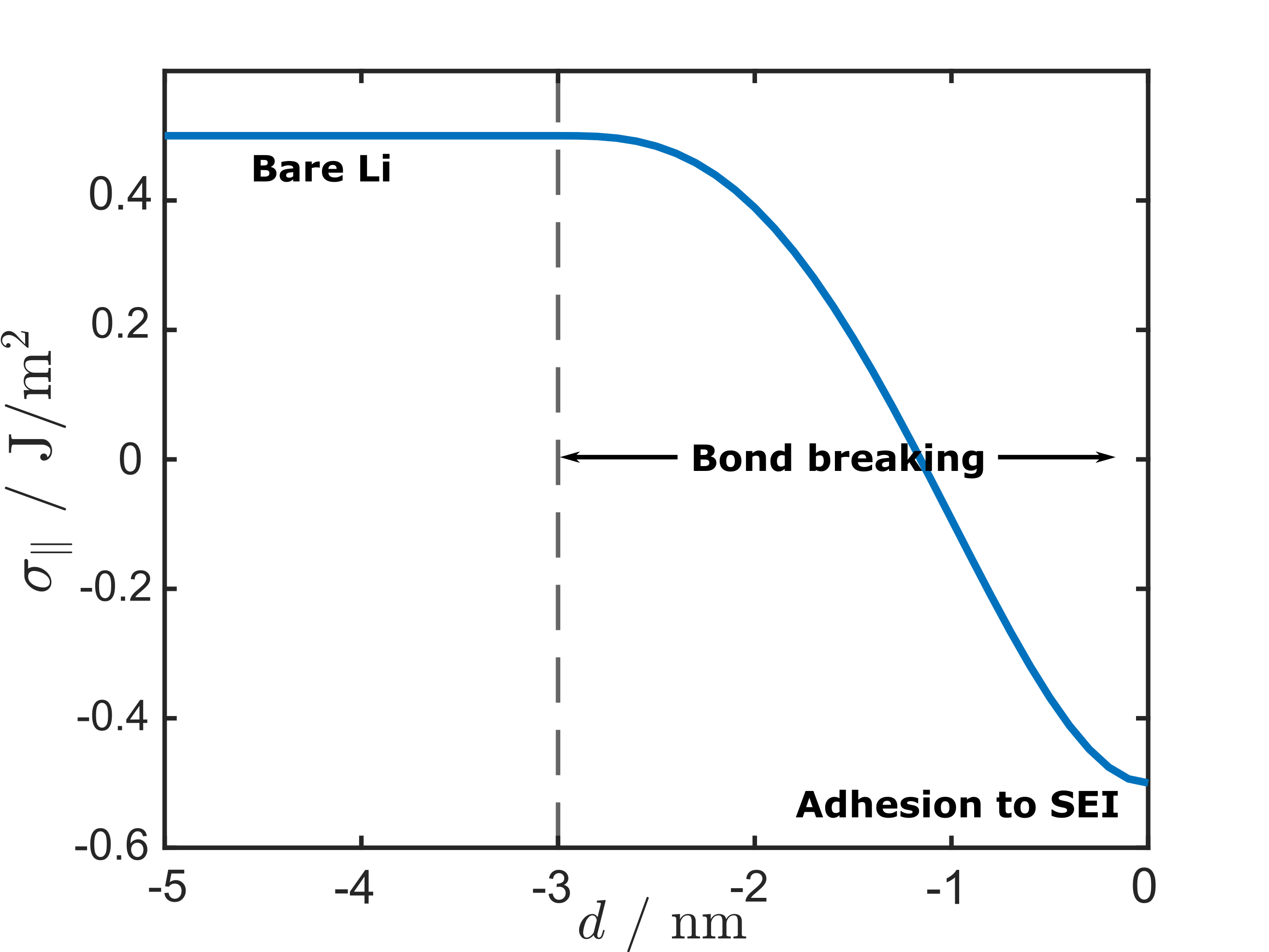}
     \caption{Change of effective interfacial energy $\sigma_\parallel$ as a function of distance $d$ to the SEI modelled by Eq.~A9.}
     \label{Fig:sigma_parallel}
\end{figure}

In order to get an expression for the chemical potential $\mu$, we model the Gibbs energy density $g$. In our model, the change of $g$ is due to the change of surface tension $\sigma$. At the beginning of the experiment, the lithium surface is parallel to the rigid SEI surface, and the lithium is bonded to the SEI. In order to strip a lithium atom from underneath the SEI, the work of adhesion has to be done. At the end of the experiment, the lithium and the SEI are decoupled. We model this with the function $\sigma_\parallel(d)$, where $d$ is the distance between lithium and the SEI. The distance dependence mimics the behavior of typical molecular-binding potentials. In our continuum approach, we further need to account for the case  that lithium can be orthogonal to the SEI, e.g., when a gap is formed in the lithium whisker. In this situation, there is no binding between lithium and the SEI. We model this with the function $\sigma_\perp=\sigma_\text{Li}$, where $\sigma_\text{Li}=\SI{0.5}{J / m^2}$ is the surface free energy of lithium~\cite{Santos2021}. We, therefore, combine both parts with an angle-dependent function $f(\alpha)$, where $\alpha$ is the angle between the SEI and the normal of the whisker surface:
\begin{equation}
\label{EQ:sigma_general}
    \sigma(d,\alpha)=\sigma_\parallel(d) f(\alpha) + \sigma_\perp (1-f(\alpha)) ,
\end{equation}
with $\sigma_\parallel(d)=\sigma_\text{Li}+E_\text{A}(d)$ including the surface free energy of lithium $\sigma_\text{Li}$ and the work of adhesion $E_\text{A}$ due to the binding to the SEI. By $\sigma_\parallel(0)=-\sigma_\text{Li}$, we assume that the bond strength from lithium to the SEI is in the same order of magnitude as the lithium-lithium cohesive bond. For our continuum approach, we smear out the bond breaking over the distance $a=\SI{3}{nm}$ so that $\sigma_\parallel(d \le -a)=\sigma_\text{Li}$. The function $\sigma_\parallel(d)$ is depicted in Fig.~\ref{Fig:sigma_parallel}. The angle-dependence is chosen such that in the perpendicular case $\sigma(d,\alpha=90^\circ)=\sigma_\perp=\sigma_\text{Li}$, i.e. $f(\alpha=90^\circ)=0$. For numerical stability, we choose $f(\alpha\le 45^\circ)=\cos(2\alpha)$ and $f(\alpha > 45^\circ)=0$. Further details are presented in the SI.

With these definitions, we can evaluate Eq.~\ref{EQ:variational_chem_pot} and calculate the chemical potential. The full derivation is presented in the SI with the final result being Eq.~A15. Initially, for $d=0$ the adhesion to the SEI leads to a negative chemical potential of the lithium surface, as $\sigma_\parallel$ is negative. This implies, that concave surfaces have a higher chemical potential, see Eqs.~A15 \& \ref{Eq:Young_Laplace}. The bond breaking leads to a steadily decreasing value of the chemical potential until $d\approx \SI{-1}{nm}$. After this the chemical potential rises again until its detached from the SEI. During the bond breaking there will be a point, at which $\sigma_\parallel$ switches its sign and concave surfaces will have a lower chemical potential. When the bond is broken, from Eq.~\ref{EQ:variational_chem_pot}, we recover the same expression for the chemical potential as by inserting the well known Young-Laplace equation in the Gibbs-Duhem relation:

\begin{equation}
    \label{Eq:Young_Laplace}
    \mu= V_\text{M} \sigma_\text{Li} \left(\frac{1}{R_1}+\frac{1}{R_2}\right) .
\end{equation}

We only model the dynamics of the lithium, not of the SEI. If, however, an instability occurs leading to droplet formation, we assume that the SEI shell breaks as the elastic deformation can only account for a few percent of volume change~\cite{Yoon2020,Kolzenberg2021}. Then, electrolyte can flood underneath the SEI shell and break the lithium-SEI bond. We assume this process to occur instantaneously, after the droplet formation. We assume that lithium is disconnected from the current collector at the point where the whisker thickness falls below the threshold $r(\xi)<0.05R$. 

When the SEI breaks, the lithium ions do not have to diffuse through the complicated SEI, and thus the effective exchange current density becomes larger. This is suggested by experiments where the exchange current density was measured. Shi \textit{et al.} used a microelectrode setup and cyclic voltammetry with a large sweeping rate of $\SI{200}{mV~s^{-1}}$ and claimed to measure the exchange current density of lithium metal deposition without SEI~\cite{Shi12138}. In 1M LiTFSI 1,3-dioxolane/1,2-dimethoxyethane electrolyte they measured $J_0=\SI{1230}{A~m^{-2}}$. For the same electrolyte, Chen \textit{et al.} measured $J_0=\SI{4.1}{A~m^{-2}}$ using lithium-lithium symmetric cells and a sweeping rate of $\SI{0.5}{mV~s^{-1}}$~\cite{Chen2020}. There, the SEI covering the lithium explains the discrepancy of these values. To the best of our knowledge, there is no systematic study on how the SEI thickness influences the effective exchange current density. Lithium whiskers are typically covered by a very thin SEI of about $\SI{20}{nm}$~\cite{Xu2020}. Therefore, we assume a relatively large exchange current density of $J_0=\SI{100}{A~m^{-2}}$ with the lithium adhered to the SEI and an exchange current density $J_0=\SI{1000}{A~m^{-2}}$ after the SEI breaks.

\subsection{Experimental Setup}
\label{Section:Experimental_Setup}

\subsubsection{Electrochemistry}
We assembled a CR2032 coin
cell with a Cu TEM grid on a Cu foil as the working electrode, lithium metal as the counter electrode and reference electrode, and polyethylene as the separator in an argon (Ar)-filled glovebox. The diameter of lithium metal was $\SI{1.56}{cm}$, and the diameter of Cu foil was around $\SI{1.8}{cm}$. A polyethylene separator was used to separate the two electrodes. The electrolyte was 1.2 M LiPF6 in ethylene carobonate (EC)/ethyl merhyl carbonate EMC (3:7 by weight) with 5 wt \% vinylene carbonate (VC). Lithium metal was deposited onto the working electrode by applying different current densities of $\SI{1}{A m^{-2}}$ for 100 minutes (by using Arbin BT-2000). After deposition, lithium was stripped by applying a current density of $\SI{-0.1}{A m^{-2}}$ for 500 minutes.

\subsubsection{Transfer to the cryo-TEM}
After cycling, the coin cell was disassembled in the Ar-filled glovebox. The TEM grid was taken out of the Cu foil and slightly rinsed with EMC to remove trace electrolyte. After rinsing, the TEM grid was placed in a sealed bag filled with Ar. Immediately after taking the sealed bag out from the Ar-filled glovebox, it was plunged into a bath of liquid nitrogen until the lithium metal reach very low temperature (around 100 K). Then, we quickly took the Cu TEM grid with electrochemically deposited lithium from the sealed bag and loaded it onto a precooled Gatan cryoholder (Elsa, Gatan, USA) using a cryotransfer station to ensure the entire process occurred under a cryogenic environment. This preserves the specimen in its native state.

\subsubsection{Cryo-TEM characterization of the lithium deposits after cycling}
A $\SI{300}{kV}$ FEI Titan monochromated (scanning) transmission electron microscope ((S)TEM) equipped with a probe aberration corrector was used to acquire the TEM, selected area electron diffraction (SAED), energy dispersive spectroscopy (EDS), and EELS data. The samples were imaged at low temperature ($\SI{100}{K}$) under low dose conditions ($\sim \SI{1}{e \per \angstrom ^{2} s}$ for low magnification imaging, $\sim \SI{100}{e \per \angstrom ^{2} s}$ for high resolution TEM imaging) to prevent beam induced damage and artifacts. EDS elemental mapping was collected by scanning the same region multiple times at a dwell time of $\numrange[range-phrase = -]{1}{10} ~ \si{\mu s}$ (depending on the image size), and the dose rate was around $\numrange[range-phrase = -]{0.363}{1.98} ~ \si{e \per \angstrom ^{2} s}$ depending on magnification. The functions of binning and smoothing in Aztec software (Oxford Instruments) were used to enhance the contrast of EDS data. Spectroscopy experiments were performed on a Gatan GIF-Quantum spectrometer. The EELS collection semi angle during the spectroscopy experiments was $\sim \SI{45}{mrad}$. EELS spectra dispersion was $\SI{0.05}{eV \per channel}$ with vertical binning at 130 in dual EELS mode. The probe beam current was around $\SI{25}{pA}$, and pixel dwell time was $\numrange[range-phrase = -]{0.001}{0.5} ~ \si{s}$. The electron dose applied during acquisition of the EELS spectra was $\numrange[range-phrase = -]{0.8}{40} ~ \si{e \per \angstrom ^{2}}$. These electron dose rates are typically used in cryogenic environment and do not introduce obvious damage or artifacts after acquiring images, diffraction patterns, EDS, and EELS spectra~\cite{Xu2020,Xu2020_2,Zachman2018,Li506,Huang2020,LI20182167}.

\begin{acknowledgement}

The work performed at Pacific Northwest National Laboratory (PNNL) was supported by the Assistant Secretary for Energy Efficiency and Renewable Energy, Office of Vehicle Technologies of the U.S. Department of Energy (DOE) und the Advanced Battery Materials Research (BMR) Program and the US-Germany Cooperation on Energy Storage with Contract Nos. DE-LC-000L072 (for C.W.) and DE-AC05-76RL01830 (for W.X.). The microscopic and spectroscopic characterizations were conducted in the William R. Wiley Environmental Molecular Sciences Laboratory (EMSL), a national scientific user facility sponsored by DOE's Office of Biological and Environmental Research and located at PNNL. PNNL is operated by Batelle for the DOE under Contract DE-AC05-76RL01830. M.W., A.L., and B.H. greatfully acknowledge financial support within the Lillint project (03XP0225A). The support of the bwHPC initiative through the use of the JUSTUS HPC facility at Ulm University is acknowledged. B.H. and A.L. acknowledge Dominik Kramer for fruitful discussions. 

\end{acknowledgement}

\section{Author's contribution}
M.W.: conceptualization, methodology, software, validation, formal analysis, data curation, writing - original draft, visualization; Y.X.: conceptualization, validation, investigation, data curation, writing - review \& editing; H.J.: conceptualization, validation, investigation, data curation, writing - review \& editing; C.W.: conceptualization, resources, writing - review \& editing, supervision, project administration, funding acquisition; W.X.: conceptualization, resources, writing - review \& editing, supervision, project administration, funding acquisition; A.L.: conceptualization, resources, writing - review \& editing, supervision, project administration, funding acquisition; B.H.: conceptualization, methodology, resources, writing - review \& editing, supervision, project administration, funding acquisition.


{ \scriptsize
\setlength{\parskip}{0pt}
\bibliography{references}
}

\end{document}